\documentclass[preprint]{elsarticle}

\begin{document}

\title{Creation of graphene allotropes using patterned defects}
\author{Mark T. Lusk*, L.D. Carr}
\ead{mlusk@mines.edu}
\address{Department of Physics, Colorado School of Mines, Golden, CO 80401, USA}
\cortext[cor1]{Corresponding author}

\begin{abstract}
Monolithic structures can be built into graphene by the addition and subsequent re-arrangement of carbon atoms. To this end, ad-dimers of carbon are a particularly attractive building block because a number of emerging technologies offer the promise of precisely placing them on carbon surfaces. In concert with the more common Stone-Thrower-Wales defect, repeating patterns can be introduced to create as yet unrealized materials. The idea of building such allotropes out of defects is new, and we demonstrate the technique by constructing two-dimensional carbon allotropes known as haeckelite. We then extend the idea to create a new class of membranic carbon allotropes that we call \emph{dimerite}, composed exclusively of ad-dimer defects.
\end{abstract}

\maketitle

\section{Introduction}
Nanoengineering seeks to create structures through direct, atom-by-atom, manipulation.
Substrate material can serve as an inert base for the addition of new species required for device
construction, but a simpler scenario is one in which the substrate itself is restructured in a way
that gives it a desirable technological property.  This is a familiar theme in traditional materials science,
where the introduction of defects modifies mechanical strength, electrical conductivity, and chemical reactivity.
Atomic level manipulation, though, provides the fidelity to treat defects as building blocks in the creation
of new structures. In this setting, the appellation of \textit{defect} is a misnomer since the disruption of
lattice periodicity is by design; plants are referred to as weeds only when they appear where they are not desired.
In this Letter, we show how a new class of carbon allotropes, called \emph{dimerites}, as well as all previously known
allotropes in 2D, can be constructed from defects.

\begin{figure}[ptb]\begin{center}
\includegraphics[width=0.45\textwidth]{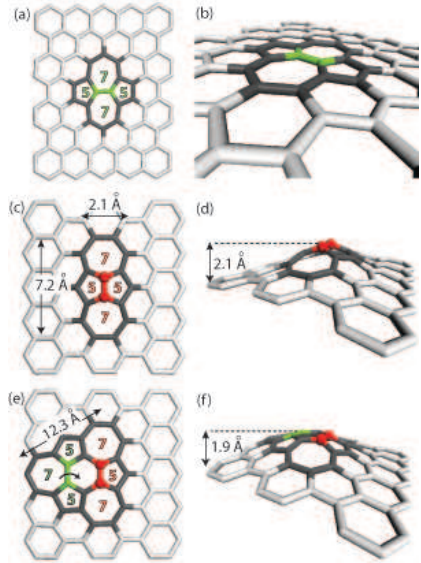}
\caption{Top and perspective views of (a-b) a Stone-Thrower-Wales (STW) defect, (c-d)
an Inverse Stone-Thrower-Wales (ISTW) defect, and (e-f) an ISTW defect which has been
partially dissociated through the addition of an STW defect.}
\label{STW_and_ISTW}
\end{center}\end{figure}
In the past, defect-based nanoengineering of carbon structures tended to focus on tubules in
order to further extend the technological reach of this novel form of matter.  The
removal or addition of carbon atoms to nanotubes has a dramatic impact on their electrical~\cite{lee2005,biercuk2004}
and mechanical properties~\cite{salvetat1999}, and can be used to weld them together\cite{terrones2002} or
break them apart~\cite{ajayan1998}.  These defects can be introduced by ion irradiation~\cite{krasheninnikov2002},
electron irradiation~\cite{smith2001,cronin2006}, and scanning tunneling microscopy (STM)~\cite{berthe2007}.
Beyond single vacancies and single adatoms, defect structures can be introduced which
involve more substantial rearrangements of the carbon lattice.  For instance, the ubiquitous Stone-Thrower-Wales (STW)
defect shown in Fig.~\ref{STW_and_ISTW}(a, b) is constructed by a simple rotation of a pair of carbon
atoms to form pairs of opposing 5- and 7-membered rings.

Another defect structure composed of opposing 5- and 7-membered rings can be fabricated by
the addition of two carbon atoms as shown in Fig.~\ref{STW_and_ISTW}(c, d). Such ad-dimer
defects were first proposed by Nardelli {\it et al.}~\cite{nardelli2000} with an eye towards creating
quantum dots from nanotube constrictions.  Molecular dynamics and tight binding simulations indicate
that these defects will dissociate under strain through the generation of Stone-Thrower-Wales
defects and can thus facilitate plastic deformation. Sufficiently large strains would then result
in defect structures that girdle the tube with the appearance of a localized constriction -- a quantum dot
along the one-dimensional conduit.  Because of their predicted influence on both mechanical
and electrical properties of carbon nanotubes, ad-dimer defects have continued to
receive attention in theoretical nanotube investigations~\cite{nardelli2000,ewels2002,sternberg2006}.
Recent progress in STM tailoring of defects on nanotubes suggests that this may be a promising route for
creating such ad-dimers defects and exploiting their technological potential~\cite{berthe2007}. Even
more precise atom-by-atom manipulation via atomic force microscopy (AFM) could also be used to
tailor defect structures~\cite{sugimoto2005}. A device has even been proposed, as a third possibility,
that is specifically designed to deliver carbon dimers to graphene- and diamond-like materials~\cite{allis2005}.

\begin{figure}[t]\begin{center}
\includegraphics[width=0.45\textwidth]{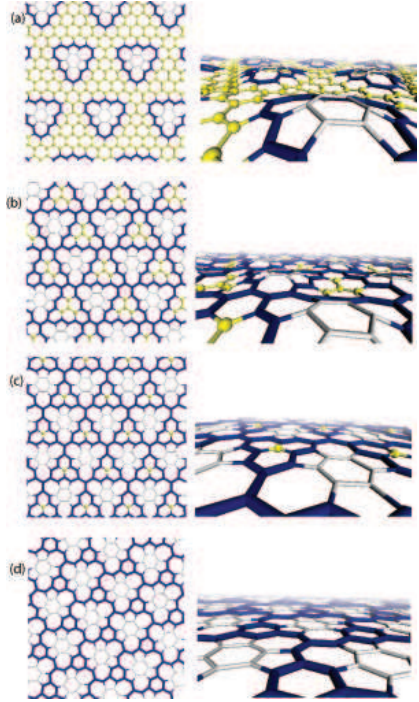}
\caption{Graphene meta-crystals, a new artificial material.  The effect of blister packing on out-of-plane distortion is shown.  As the amount of interstitial graphene (yellow) increases, so does the out-of-plane distortion. It is the incompatibility of the allotropes that causes this distortion.  (a)-(d) Top and side perspective views of metacrystals. The primitive cells are, respectively, a rhombus ($12.32$\AA, $120^{\circ}$),  a parallelogram ($7.63$\AA $\times 9.20$\AA, $73.3^{\circ}$), a rhombus ($7.97$\AA, $120^{\circ}$), and a rhombus ($7.08$\AA, $120^{\circ}$). The meta-crystals all exhibit six-fold symmetry with meta-atom spacings of $12.32$\AA, $10.13$\AA, $7.99$\AA, and $7.08$\AA, respectively.}
\label{blister_packing}
\end{center}\end{figure}
The recent experimental realization of graphene~\cite{novoselov2005,zhangYB2005} brings the consideration of
graphitic nanoengineering to a new plane. These single-layer carbon structures have been the subject of
significant experimental and theoretical queries because of their potential within the electronics
industry and because they represent a new and poorly understand form of matter. Just like their
tubule counterparts, graphene sheets are known to have defects~\cite{hashimoto_2004,stroscio2007a,stroscio2007b}.
At the most basic level, defects can be formed by knocking atoms out of the lattice so as to have
vacancies of various sizes; alternately, surplus atoms can be found as adatoms on the surface.
Ad-dimer defects can also be introduced into graphene and modified via STW defects as we demonstrated in recent
work~\cite{LuskCarr2008a}.  Within this setting, the defects appear as protrusions from the graphene plane
as shown in Fig.~\ref{STW_and_ISTW}(c, d).  Because the ad-dimer defect has adjacent 5-membered rings
instead of adjacent 7-membered rings, it was dubbed an \textit{Inverse Stone-Thrower-Wales (ISTW) defect}.
The introduction of a STW defect causes the ISTW defect to dissociate into a {\em ISTW-STW blister}
with three-fold symmetry as shown in Fig.~\ref{STW_and_ISTW}(e, f). Taken together, the STW and ISTW
defects may be viewed as basic building blocks in a taxonomy of structures with novel thermo-mechanical,
chemical,  electrical, and magnetic properties.

These graphene defects can be patterned so as to synthesize entirely new forms of carbon membranes.
Templated arrangements can give graphene the appearance of bubble wrap and is suggestive of
two-dimensional metacrystals with defect meta-atoms.  This perspective facilitates the application of
group symmetry concepts and density functional theory (DFT) to design both planar and non-planar
carbon membranes. Specifically, the ground state of promising arrangements of defects can be used
to identify low energy allotropes that could be experimentally synthesized. This structural focus
is a pre-requisite to exploring electronic and other properties of defects on graphene
composed entirely of carbon atoms.

\section{Theory}
We first consider an isolated ISTW defect by placing a single ad-dimer above opposing bridge
sites on a hexagonal ring of carbon atoms. These adatoms interact with the graphene to form a backbone
for adjacent 5-membered rings as shown in figure Fig.~\ref{STW_and_ISTW}(c, d).  The defect involves the
re-structuring of three hexagonal rings laid out in a row, so that the final ISTW defect structure is elongated,
as can be observed in the figure. All carbon atoms still covalently bond with three nearest neighbors and so
have unbonded $\pi_z$ electrons out of plane which give the graphene much of its interesting charge
transport character.

Although ISTW defects have yet to be observed, certain experiments have
provided tantalizing hints~\cite{heben1991,stroscio2007a, stroscio2007b}.  Moreover, DFT analysis indicates that
ISTW defects may well be found on the lip of divacancies~\cite{LuskCarr2008a}. In fact, in graphene systems with a
sufficient number of carbon adatoms, ISTW defects might also be found away from vacancies because
the barrier required to form them from the chance meeting of two adatoms is relatively low.
It is therefore reasonable to search for such structures on graphene as a first step towards their
experimental elucidation.  As a second step towards nanoengineering with ISTW defects, electron/ion
irradiation~\cite{krasheninnikov2002,smith2001,cronin2006} or di-carbon molecular bombardment could
be used to generate such defect structures without concern about their relative placement or type.
These activities may well uncover blisters composed of combinations of STW and ISTW defects.

Having considered isolated defects, we now turn to the prospect of patterning them on graphene. Templated ISTW-STW blisters are used to introduce the concept. The periodic cells associated with Fig.~\ref{STW_and_ISTW} may be
viewed as {\em metacrystals} with each defect representing a {\em meta-atom}. Of particular interest, though,
are systems in which the meta-atoms are sufficiently close to exhibit meaningful interaction.
For instance, Fig.~\ref{blister_packing}(a) shows the ground state of a network of
ISTW-STW blisters separated by a contiguous graphene lattice.
The cell geometry has been optimized, and an out-of-plane distortion is clearly evident.
The degree of nonplanarity correlates with the degree to which the {\em interstitial} graphene
resists the stress imposed on it by the blisters. As the amount of graphene is reduced, the
out-of-plane distortion reduces (Fig.~\ref{blister_packing}(b, c)). A close-packed arrangement of the
blister meta-atoms, wherein no interstitial graphene is present, results in a planar
structure known as H$_{5,6,7}$ haeckelite (Fig.~\ref{blister_packing}(d))~\cite{terrones2000}.  Haeckelites have yet to be
experimentally realized, and using defects to create patches of them may provide a reasonable synthesis strategy.
Interestingly, such patches would have a phase interface with the graphene, and the accretive mobility
and kinetics of such boundaries is untouched territory.

\begin{figure}[t]\begin{center}
\includegraphics[width=0.45\textwidth]{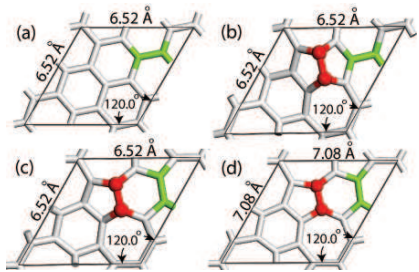}
\caption{H$_{5,6,7}$ haeckelite synthesis from graphene. (a) graphene lattice with no defects but with bond of future STW defect highlighted in green. (b) ISTW defect is introduced (red). (c) STW defect is introduced by rotating a single bond (green). (d) lattice is allowed to dilate and the result is a planar carbon allotrope.}
\label{graphene_to_H567}
\end{center}\end{figure}
In order to elucidate a procedure for synthesizing new allotropes, we now show how this form of
haeckelite can be constructed explicitly from patterned ISTW-STW blisters arranged on clean graphene (Fig.~\ref{graphene_to_H567}).
Blister formation within a periodic graphene cell of appropriate dimensions results in a precursor
to H$_{5,6,7}$ haeckelite.  Without changing the cell size, a meta-crystal is formed wherein the blisters
exhibit a clear out-of-plane distortion~\cite{LuskCarr2008a}. However, an optimization of the cell size
results in the planar haeckelite structure shown in Fig.~\ref{graphene_to_H567}(c, d).  Planar
haeckelite requires that the side length expand from $6.52$\AA  to $7.08$\AA, a linear dilation of 8.6\% .
We found a ground state energy estimate for H$_{5,6,7}$ of 0.23 eV/atom above graphene.  This is
close to the literature value of 0.25 eV/atom~\cite{terrones2004} and notably lower than that of
0.38 meV/atom for C$_{60}$~\cite{LuskCarr2008a}. Interestingly, both O$_{5,6,7}$ and R$_{5,7}$
forms of haeckelite can be formed using only STW defects.  However, these are of somewhat higher energy.

\begin{figure}[t]\begin{center}
\includegraphics[width=0.45\textwidth]{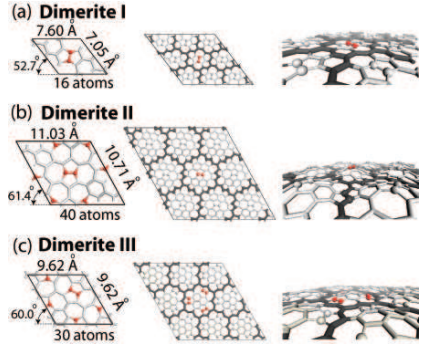}
\caption{Dimerites, new allotropes of carbon in two dimensions which can be constructed
purely of ISTW defects. Three such structures are shown. The primitive cells are given in column one and views with nine primitive cells are provided in columns two (top view) and three (perspective view). The darkened lines in columns two and three highlight the borders of meta-atoms. Red atoms are associated with carbon dimers.}
\label{dimerite_all}
\end{center}\end{figure}
ISTW-STW blisters, used to introduce meta-atoms, require both dimer additions and bond rotations.
It is likely, though, that it would be easier to synthesize crystals which to not require STW defects
at all and are based exclusively on ISTW defects. We refer to such materials as \emph{dimerites}.
Strictly speaking, the periodic cells associated with Figs.~\ref{STW_and_ISTW}(c, d)
are members of this class, but our intent is introduce materials with closer meta-atom packing.
Three such metacrystals are shown in Fig.~\ref{dimerite_all}.  Dimerite I is planar and is the lowest
energy configuration (Fig.~\ref{dimerite_all}(a))
with a ground state energy 0.28 eV/atom above graphene. This is only 0.04 eV/atom higher
than H$_{5,6,7}$ haeckelite and is 0.11 eV/atom lower than C$_{60}$.  The ISTW defects have been
arranged so as to cancel the elastic strains which result in out-of-plane distortion.  Dimerites II and III (Fig.~\ref{dimerite_all}(b, c) are non-planar and exhibit higher symmetry at the expense of higher ground
state energies: 0.33 eV/atom and 0.37 eV/atom above graphene, respectively.
Dimerite II is particularly intriguing because of the six-fold symmetry of the meta-atom perimeter and three-fold internal symmetry. The symmetry groups of the three allotropes are P2/M, P2, and P31M, respectively.

\begin{figure}[ptb]\begin{center}
\includegraphics[width=0.45\textwidth]{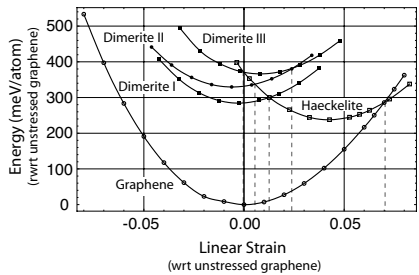}
\caption{Mechanical response (pure dilation) of graphene, H$_{5,6,7}$ haeckelite, and three dimerites. The points represent actual DFT data; the curves are a
guide to the eye. Dashed vertical lines indicate strains at which there is a stability exchange between carbon allotropes.}
\label{dimerite_elasticity}
\end{center}\end{figure}

As is clear in Fig.~\ref{dimerite_elasticity}, dimerites I and II are slightly more dense than the 0.38 atoms/\AA$^2$  of
graphene (+0.5$\%$ and +1.2$\%$). Dimerite III is 1.6$\%$ less dense than graphene.  The uniaxial elastic constant ($C_{11}$) of dimerite I is 14.4$\%$ more than that of graphene (1.13 GPa), while dimerites II and III are 10.4$\%$ and 6.0$\%$ less stiff than graphene, respectively. The softer tensile nature of the latter two allotropes is attributable to the flattening of out-of-plane undulations in response to loading. Fig.~\ref{dimerite_elasticity} also shows the mechanical response of H$_{5,6,7}$ haeckelite~\cite{LuskCarr2008a}. In comparison with this allotrope, the dimerites all have ground state densities much closer to graphene, implying lower mismatch strains for forming dimerites patches and strips within graphene. The plot also indicates three strains at which a stability exchange occurs between a dimerite and the haeckelite. A fourth stability exchange is shown between dimerites II and III. The existence of these critical points indicate that phase transitions can be induced within carbon sheets through the application of modest strains.

\section{Conclusion}
In conclusion, we have presented a new class of carbon allotropes in two dimensions which we call
\emph{dimerite}.  We showed that these and other 2D carbon allotropes can be nanoengineered directly from graphene
via two types of defects.  Dimerites in particular can be made solely by adding pairs of atoms to the
surface of graphene. So far as we know, there are no dimerites which occur spontaneously in Nature, and defect engineering even small dimerite patches
 will require precise atomic control. Recent technological advances, though,
allow the conclusion that the requisite fidelity is now experimentally accessible.
Scanning tunneling microscopy~\cite{berthe2007}, atomic force microscopy~\cite{sugimoto2005}, and devices of the sort described by Allis and Drexler~\cite{allis2005} offer the brightest prospects for such graphene nanoengineering. It might also be possible to carry out metacrystal construction using interference patterns of electrons in the near field of a diffraction grating~\cite{cronin2006}.  The approach has the
potential to create features with a periodicity on the scale of 10 nm.

\section{Acknowledgement}
LDC was supported by the National Science Foundation under Grant PHY-0547845 as part of the NSF Career program.
We thank David Wu for useful conversations.


\end{document}